
%
%

 \input harvmac

\def\const {{\rm const}}
\def \s {\sigma}
\def\t {\tau}

\def \ha {\half}
\def \ov {\over}
\def \fourth  {{1\ov 4} }
\def \four{{\textstyle {1\ov 4}}}
\def \a {\alpha}
\def \lr { \lref}
\def\ep{\epsilon}

\def \bd {\bar \del}
\def \r {\rho}
\def\const {{\rm const}}\def\bd {\bar \del} \def\m{\mu}\def\n {\nu}\def\l
{\lambda}

\def\y {{ \tilde y}}

 \def \sm {$\s$-model\ }
\def   \td {\tilde }

\def \lr { \lref}

\gdef \jnl#1, #2, #3, 1#4#5#6{ { #1~}{ #2} (1#4#5#6) #3}

\lr \orb {L. Dixon, J. Harvey, C. Vafa and E. Witten, \np B274 (1986) 285.}

\lr \burg{C.P. Burgess, \np B294 (1987) 427; V.V. Nesterenko, \ijmp A4 (1989)
2627.}
\lr \tsenul {A.A. Tseytlin, \np B390 (1993) 153. }
\lr \susskind { L. Susskind, ``Some speculations about black hole entropy in
string theory", RU-93-44 (1993), hep-th/9309135. }
 \lr \dabh {A. Dabholkar, ``Strings on a cone and black hole entropy",
HUTP-94-A019,  hep-th/9408098; ``Quantum corrections to black hole entropy in
string theory",    hep-th/9409158;
D.A. Lowe and A. Strominger, ``Strings near a Rindler or black hole
horizon", hep-th/9410215.}

\lr\hrt {G.T. Horowitz and A.A. Tseytlin. \pr D50 (1994) 5204. }

\lr \gtwo {E. Del Giudice, P. Di Vecchia and S. Fubini, Ann. Phys. 70
(1972) 378; K. A. Friedman and C. Rosenzweig, Nuovo Cimento 10A (1972) 53;
S. Matsuda and T. Saido, Phys. Lett. B43 (1973) 123; M. Ademollo {\it et al},
Nuovo Cimento A21 (1974) 77;
S. Ferrara, M. Porrati and V.L. Teledgi, Phys. Rev. D46
(1992) 3529.}

\lr \plane { D. Amati and C. Klim\v c\'\i k,
\jnl \pl, B219, 443, 1989;
 G. Horowitz and A. Steif,  \jnl \prl, 64, 260, 1990; \jnl \pr,
D42, 1950, 1990;
 G. Horowitz, in: {\it
 Strings '90}, eds. R Arnowitt et. al.
 (World Scientific, Singapore, 1991);
 H. de Vega and N. S\' anchez, \pr D45 (1992) 2783; Class. Quant. Grav. 10
(1993) 2007.}

\lr \gibma { G.W.  Gibbons and  K. Maeda, \np B298 (1988) 741;
G.W.  Gibbons, in: {\it Fields and Geometry}, Proceedings of the 22-nd Karpacz
Winter School of Theoretical Physics, ed. A. Jadczyk (World Scientific,
Singapore,  1986).}

\lref \tsnul { A. Tseytlin, \jnl \np, B390, 153, 1993.}

\lr\duval  { C. Duval, Z. Horvath and P.A. Horvathy, \jnl \pl,  B313, 10,
1993.}

\lr\gro{D.J.  Gross, J.A. Harvey,  E. Martinec and R. Rohm, \np B256 (1985)
253; \np B267 (1985) 75.}
\lr \duff {Duff et al Nepomechie }
\lr \incer { E.J. Ferrer, E.S. Fradkin and V.de la Incera, \pl B248 (1990)
281.}
\lr \green {M.B. Green, J.H.  Schwarz and E.  Witten, {\it Superstring Theory}
(Cambridge U.P., 1988).}

\lr \quev {C. Burgess and F. Quevedo,  \np B 421 (1994) 373. }
  \lr \nahm { W. Nahm, \np B124 (1977) 121. }
 \lr \kkkk {  E. Kiritsis and C. Kounnas,  ``Curved four-dimensional spacetime
as infrared regulator in superstring theories", hep-th/9410212. }
\lr \sen{A. Sen, \pr D32 (1985) 2102; \prl 55 (1985) 1846.}
\lr \hulw { C. M. Hull and E.  Witten,  \jnl \pl, B160, 398, 1985. }

\lr\hult { C. Hull and P. Townsend, \jnl \pl, B178, 187, 1986. }
\lr \gps {S.  Giddings, J. Polchinski and A. Strominger, \jnl  \pr,  D48,
 5784, 1993. }

\lr \ghr { S. Gates, C. Hull and M. Ro\v cek, \np B248 (1984) 15.}
\lr \jon {C. Johnson, \pr D50 (1994) 4032.}
\lr \landau {L.D.  Landau and E.M.  Lifshitz, {\it Quantum Mechanics} (Pergamon
Press, N.Y., 1977).  }
\lref \horts {G.T. Horowitz and A.A. Tseytlin, \pr  D51 (1995) 2896. }

\lr \attick {J.J. Attick  and E. Witten, \np B310 (1988) 291. }
\lr\fund{A. Dabholkar, G. Gibbons, J. Harvey and F. Ruiz Ruiz, \jnl \np, B340,
33, 1990;
D. Garfinkle, \jnl \pr, D46, 4286, 1992; A. Sen, \jnl \np, B388, 457, 1992;
D. Waldram, \jnl \pr, D47, 2528, 1993. }

\lr\planetach{ J. Garriga and E. Verdaguer, Phys. Rev. {\bf D43} (1991) 391.}

\lr \duality { L. Brink, M. Green and J. Schwarz, \np B198 (1982) 474;
K. Kikkawa and M. Yamasaki, \pl B149 (1984) 357;
N. Sakai and I. Senda, { Progr. Theor. Phys. }  75 (1984) 692. }

\lr \canon {A. Giveon, E. Rabinovici and G. Veneziano, \np B332 (1989)167;
K. Meissner and G. Veneziano, \pl B267 (1991) 33;
E. \' Alvarez, L. \' Alvarez-Gaum\' e and Y. Lozano, \pl B336 (1994) 183. }

\lr \kkl {E. Kiritsis, C. Kounnas and D. L\" ust, \pl B331 (1994) 321.}
\lr \kk {E. Kiritsis and  C. Kounnas,  \pl B320 (1994) 361.}
\lref \kallosh { E. Bergshoeff, R. Kallosh and T. Ort\' \i n, \jnl \pr,  D47,
5444,
1993; E. Bergshoeff, I. Entrop and R. Kallosh,
\jnl \pr, D49, 6663, 1994.   }

\lr\tseta{A.A. Tseytlin, \pl B317 (1993) 563;
K. Sfetsos and A.A. Tseytlin, \pr D49 (1994) 2933.  }

\lr\frats { E.S. Fradkin and A.A. Tseytlin , \pl B163 (1985) 123. }
\lr\tsey{  A.A. Tseytlin, \pl B202 (1988) 81.  }
\lr\tset { A.A. Tseytlin, \np B350 (1991) 395.}
\lr\metsa {R.R. Metsaev and A.A. Tseytlin, \np B298 (1988) 109. }
\lr\abo { A. Abouelsaood, C. Callan, C. Nappi and S. Yost, \np B280 (1987) 599.
}
\lr\bachas {C. Bachas and M. Porrati , \pl B296 (1992) 77.}
\lr\ferrara{S. Ferrara and M. Porrati, \mpl A8 (1993) 2497.}
\lr \sfets { K. Sfetsos and  A.A. Tseytlin, \np B427 (1994) 325.  }
\lr\rs {J.G. Russo and L. Susskind,   \np B437 (1995) 611.}
\lr\thermal {J.G. Russo, \pl B335 (1994) 168.}
\lr \napwi{ C. Nappi and E. Witten, \jnl \prl, 71, 3751, 1993.}

\lr \narain{K. Narain, \pl B169 (1986) 41;
K. Narain, H. Sarmadi and E. Witten, \np B279 (1987) 369;
  P. Ginsparg, \pr D35 (1987) 648.  }
\lr \sche{W. Lerche, A.N. Schellekens and N.P. Warner, Phys. Rep. 177 (1989)
1.}

\lr\tsetl { A.A. Tseytlin, \pl B208 (1988) 221.}
\lr\polch{ J. Polchinski, Commun. Math. Phys. 104 (1986) 37.}
\lr\obrien{ O'Brien -Tan , McClain-Roth }
\lr \ginsp { W. Lerche, A. Schellekens and N. Warner, Phys. Repts. 177 (1989)
1; P. Ginsparg, in: {\it Fields, Strings and Critical Phenomena, } ed. by E.
Brezin and J. Zinn-Justin (Elsevier Science Publ., 1989).  }
\lr\salam{A. Salam and J. Strathdee, \np B90 (1975) 203.}
\lr\nielsen{ N.K. Nielsen and P. Olesen, \np B144 (1978) 376;
J. Ambjorn and P. Olesen, \np B315 (1989) 606; \np B 330 (1990) 193.}

\lr\anto{I. Antoniadis and N. Obers, \jnl \np, B423, 639, 1994}
\lr \nonsemi { D. Olive, E. Rabinovici and A. Schwimmer,  \jnl \pl, B321, 361,
1994.}
\lr\sfee{
 K. Sfetsos,  \jnl \pl, B324, 335, 1994; \pr D50 (1994) 2784.}
\lr\nons{A.A.  Kehagias and P.A.A.  Meesen, \pl B331 (1994) 77;
J.M. Figueroa-O'Farril and S. Stanciu, \pl B327 (1994) 40;
A. Kehagias, ``All WZW models in $D\leq 5$", hep-th/9406136. }

\lr \busc {T.H. Buscher, \pl  B194 (1987) 59; \pl B201 (1988) 466.}
 \lr\rocver{ M. Ro\v cek and E. Verlinde, \np B373 (1992) 630. }

\lr\rut{J.G. Russo and A.A. Tseytlin, {``Exactly solvable string models
of curved space-time backgrounds"}, CERN-TH/95-20,
hep-th/9502038.}
\lr\melvint{A.A. Tseytlin,  \pl B346 (1995) 55. }
\lr\rts{J.G. Russo and A.A. Tseytlin, {``Constant magnetic field
in closed string theory: an exactly solvable   model"},
  CERN-TH/7494/94,
hep-th/9411099.}
 \lr \gross {D.J.  Gross, J.A. Harvey,  E. Martinec and R. Rohm, \np B256
(1985)
253; \np B267 (1985) 75.}
\lr \magn{I. Antoniadis, C. Bachas and A. Sagnotti, \pl B235 (1990) 255;
 J. Harvey and J. Liu, \pl B268 (1991) 40;
R. Khuri,  \pl B294 (1992) 325; \np B387 (1992)  315; J. Gauntlett, J. Harvey
and J. Liu,
 \np  B409 (1993)  363;  S.  Giddings, J. Polchinski and A. Strominger,  \pr
D48
 (1993)  5784;
C. Johnson,  \pr D50 (1994)  4032:
C. Bachas and E. Kiritsis, \pl B325 (1994) 103. }

\lr\koukir{E. Kiritsis and C. Kounnas,  ``Curved four-dimensional spacetime
as infrared regulator in superstring theories", hep-th/9410212.}

\lr\baa{ C. Bachas, ``A way to break supersymmetry", hep-th/9503030.}

\lr \gso{F. Gliozzi, J. Scherk and D. Olive, \np B122 (1977) 253;
M.B. Green, J.H.  Schwarz and E. Witten, ``Superstring Theory" (Cambridge U.P.,
1987). }

\lr \sen{A. Sen, \pr D32 (1985) 2102; \prl 55 (1985) 1846.}
\lr \hulw { C. M. Hull and E.  Witten,  \jnl \pl, B160, 398, 1985. }

\lref \kallosh { E. Bergshoeff, R. Kallosh and T. Ort\' \i n, \jnl \pr,  D47,
5444,
1993; E. Bergshoeff, I. Entrop and R. Kallosh,
\jnl \pr, D49, 6663, 1994.   }
\lr \flo{R. Floreanini and R. Jackiw, \prl 59 (1987) 1873.  }
\lr \tse { A.A. Tseytlin, \pl B242 (1990) 163; \np B350 (1991) 395.  }
\lr \ses{ A. Sen  and J.H. Schwarz, \pl B312 (1993) 105. }

\def\np {  Nucl. Phys. }
\def \pl { Phys. Lett. }
\def \mpl { Mod. Phys. Lett. }
\def \prl { Phys. Rev. Lett. }
\def \pr  { Phys. Rev. }

\def \ijmp {Int. J. Mod. Phys. }

\baselineskip8pt
\Title{\vbox
{\baselineskip 6pt{\hbox{CERN-TH/95-106}}{\hbox
{Imperial/TP/94-95/29}}{\hbox{hep-th/9506071}} {\hbox{
   }}} }
{\vbox{\centerline {Heterotic strings in  a uniform   magnetic field}
}}

\vskip -20 true pt

\centerline  { {J.G. Russo\footnote {$^*$} {e-mail address:
 jrusso@vxcern.cern.ch
} }}

 \smallskip \smallskip

\centerline{\it  Theory Division, CERN}
\smallskip

\centerline{\it  CH-1211  Geneva 23, Switzerland}

\medskip
\centerline {and}
\medskip
\centerline{   A.A. Tseytlin\footnote{$^{\star}$}{\baselineskip8pt
e-mail address: tseytlin@ic.ac.uk}\footnote{$^{\dagger}$}{\baselineskip8pt
On leave  from Lebedev  Physics
Institute, Moscow.} }

\smallskip\smallskip
\centerline {\it  Theoretical Physics Group, Blackett Laboratory}
\smallskip

\centerline {\it  Imperial College,  London SW7 2BZ, U.K. }
\bigskip\bigskip\bigskip
\centerline {\bf Abstract}
\medskip
\baselineskip8pt
\noindent
 An exact conformal  model  representing a
 constant magnetic field background in heterotic string theory is explicitly
solved in terms of free creation/annihilation operators.
The  spectrum of physical states is examined for different
 possible embeddings
of the magnetic  $U(1)$ subgroup.
We find that  an arbitrarily small magnetic field gives rise to an infinite
number of tachyonic excitations
 corresponding   to charged vector states of the massless level and  to higher
level states with large spins and charges.

\Date {June 1995}

\noblackbox
\baselineskip 14pt plus 2pt minus 2pt

\vfill\eject
 \newsec{Introduction}
 Recently,  a
 class of conformal models representing   $D=4$ axially symmetric magnetic
field backgrounds in  closed bosonic  string theory was  shown to be
exactly solvable \refs{\rts,\ \rut}. Like string theory in flat space, or
on orbifold \orb, or  open string theory in constant magnetic field \abo ,
 these quantum string  models  can be  represented  in terms of
free creation/annihilation operators, so that
 the physical spectrum,   partition function, vertex operators, etc.,
can be  explicitly  determined.
In contrast to some other solvable models, here
the underlying {\it space-time}  geometry is non-trivial (e.g. the curvature
is non-zero and may be singular in some cases).
 These backgrounds generalize exact string solutions found in \refs{\horts ,
\melvint }.

The physical spectrum has exhibited the presence of tachyonic instabilities for
{\it infinitesimal} values of the magnetic field.
 Instabilities at finite values of magnetic fields (observed  previously in
Yang-Mills gauge theory \nielsen\  and in open string theory \refs{\abo,
\ferrara })
indicate the presence of a phase transition \nielsen.\foot{Indeed,   the
thermodynamical partition function of a  string gas cannot be defined beyond
the Hagedorn temperature and beyond the critical magnetic field where tachyons
appear in the spectrum
\thermal.  It  was argued in  \thermal\  that, like  the Hagedorn transition at
zero field \attick, this is  a first-order phase transition with a large latent
heat.}
  In the case of unbroken gauge theory  the constant magnetic background
 is  unstable  already  for an infinitesimal magnetic field
 \nielsen\ (this infinitesimal instability goes away once
the gauge symmetry is spontaneously broken;   the  magnetic field  which is
necessary to
produce an instability  is then
 of the order of the mass of the charged vector bosons).

Such  infinitesimal instability is expected for charged massless
string states (members of  Yang-Mills multiplet)\foot{It was absent
in the open  (super)string models considered in \refs{\abo,  \ferrara }
since there the Chan-Paton symmetry was assumed to be Abelian
but should  of course appear in the non-Abelian case.}
and, similarly,  should disappear  after these states acquire some masses
through spontaneous symmetry breaking.
Although higher-spin   string  states may seem to be  protected from this
instability by large masses,
it nevertheless turns out that an infinite number of them become tachyonic
when an infinitesimal magnetic field is turned on \rts .
This does not happen
 in the case of the open string theory \refs{\abo, \ferrara},  where, as in the
broken gauge theory,
one needs a finite (Planck-order) magnetic field in order to  make
originally positive (mass)$^2$ string states tachyonic.
The important feature of  closed string theory is that, in contrast to the open
string case, here the charges  of states are not fixed
but (like masses and spins)  can take arbitrarily large values.
As a result  \rts,  there  are states for  which the  gyromagnetic  coupling
term ($\sim fQJ$) overpowers the  free string mass term
for a magnetic field  $f \sim 1/Q$,  which can thus be   arbitrarily  small
for large enough charge $Q$.

The  question  that will be addressed in the present paper
is whether these instabilities  appear also in the
 heterotic string  case \gross.
We shall show  that     the magnetic field  necessary to produce tachyonic
states  in the  heterotic string models is indeed
arbitrarily small.  In addition to the tachyonic charged vector modes  of the
`massless' level
there is an    infinite  number of
  tachyons  corresponding to  higher spin and  charge states of the free string
theory.\foot{This conclusion remains   valid  even if one introduces small mass
corrections
$\sim M_s<<M_{\rm Planck}$ (e.g. originating  from
supersymmetry or  gauge symmetry  breaking)  to the masses of  all  string
states.}

The heterotic string models  discussed below which  describe constant magnetic
background
were already introduced in \rts.\foot{Other magnetic (monopole-type) solutions
in heterotic string theory
were discussed, e.g., in \refs{\magn, \koukir, \baa }.}
Starting with 10-dimensional heterotic string
theory one can  embed the Abelian magnetic field
either in the Kaluza-Klein sector (assuming that 6 dimensions are compactified
on a torus,   one of the  periodic coordinates being used to couple the
magnetic field) or in the internal  $E_8\times E_8$ or $SO(32)$ gauge sector.
The two heterotic models realizing these two options
 will be discussed  below. They appear to be closely related and have similar
properties.

The Kaluza-Klein (KK)  embedding option is the only one available in the
bosonic string  and closed  superstring cases \rts.
The type II superstring model based on
direct $(1,1)$ supersymmetric generalization of the bosonic model of \rts\
turns out to have  residual space-time supersymmetry and thus no tachyons in
its spectrum.\foot{In this paper we shall use the  fermionic string (NSR)
formalism. The models we consider  can be solved also using   the Green-Schwarz
approach: the exact conformal invariance of the bosonic background
implies the existence of $\kappa$-supersymmetry and the existence
of the covariantly constant null Killing vector makes
it straightforward to fix the light-cone gauge both  for  bosons and
space-time fermions.}
   The same applies to its `left' $(1,0)$  truncation: the
corresponding heterotic string model  is stable.  It should be noted that the
`magnetic'  interpretation
of these models is rather artificial, since the
Abelian KK gauge field here  cannot be identified with the  usual
Maxwell field.

The two `right' $(0,1)$ heterotic models (with KK and  with gauge  sector
embedding) have no residual space-time supersymmetry
and exhibit   tachyonic instabilities.
Tachyonic instabilities in the presence of  an {\it infinitesimal}  magnetic
field are inevitable in any theory   containing   massless Yang-Mills vector
bosons with non-zero $U(1)_{\rm em}$ charges. What is new in  closed string
theory is that
these infinitesimal instabilities are associated also with higher level string
states and  should thus survive even after
gauge vector bosons  become  massive.\foot{ The instability could only be
removed by        Planck-order mass corrections  to massive states (for a
further discussion see Section  5).}

We shall start in  Section 2   by describing the actions of
 the heterotic models associated with a uniform magnetic background
of \refs{\rts, \horts}. We shall present  the corresponding  actions
both in the  manifestly Lorentz-invariant and in chiral boson forms,
  the latter  being useful  for establishing the relation between
the two  `right' heterotic models (which  can be interpreted as two
special cases of
the $O(6,22)$ duality-invariant  action \ses\ of toroidally compactified
heterotic string).

To find the spectra of states of these
models we shall follow the same method as
used in the bosonic case in \rts: solving explicitly the
classical equations,  quantizing the theory canonically and expressing  the
quantum Virasoro constraints in terms of free  creation/annihilation operators.
We shall first  consider  the `right'
heterotic model with KK embedding (Section 3)
and demonstrate the presence of  tachyonic instabilities
in its spectrum.  We shall also  explain why these instabilities are absent in
type II  superstring, `left-right symmetric'
and `left' heterotic string models with KK embedding
(in agreement with space-time supersymmetry of these models).

Using the results of Section 3 we shall finally determine in Section 4  the
spectrum of the  `realistic' heterotic string model with
magnetic $U(1)$ subgroup
embedded in the  $E_8\times E_8$ or $SO(32)$ internal gauge symmetry group.
As in the  case of the `right' heterotic model with KK embedding,
there is an infinite number of tachyonic states for any given arbitrarily small
value of the magnetic field strength.

Section 5 will contain a summary and  concluding remarks.

\newsec{Actions of  the heterotic string models}

\def \hv {{\hat v} }

As discussed above, our  aim will be to  solve the superstring and heterotic
string versions
of the bosonic constant magnetic field
model studied  in \rts. To embed an Abelian magnetic field in a closed
superstring theory one is to consider a toroidal compactification
(``Kaluza-Klein"  embedding).
In addition to KK embedding,  in the heterotic  string theory
there is also an option  to interpret the magnetic field as belonging to an
Abelian  subgroup of an internal gauge group (gauge sector  embedding).
The  models we shall discuss below are thus type II closed superstring
with KK embedding (and closely related `left-right symmetric' heterotic model),
its two inequivalent `left' and `right' heterotic
truncations
and the `right'  heterotic model with gauge sector  embedding of the  magnetic
field.
Many technical details of the solution of these models
will be similar.

The exact conformal invariance of these  models   as well  as their
space-time interpretation   were already discussed in  \refs{\horts,\rts}.
The corresponding  4-dimensional
space-time background  which solves the  heterotic string (as well as
compactified $D=5$ superstring and bosonic string)  equations of motion, in
particular ($\m,\n=0,1,2,3$)
\eqn\het{  R_{\m\n}-  \fourth{  H}_{\m\l\r} { H}_\n^{\ \l \r}  - \fourth \a'
{\cal F}(V)_{\m\l}  {\cal F}(V)_{\n}^{\ \  \l} + 2D_\m D_\n \Phi + ... =0  \  ,
}
is given by \horts\ (we list only the non-vanishing components of the
fields)\foot{Performing the electromagnetic duality on this background
one can find its $S$-dual, which will have non-constant dilaton (note that  the
axion is non-trivial here) and only {\it electric} component of the vector
field. The resulting string model will  not, however,  be  conformal to all
orders in $\a'$ ($S$-duality  may be
  expected to be a symmetry of the $D=4$ heterotic string only
non-perturbatively in string coupling).}
\eqn\did{ ds^2_4= -  [ dt +  A_i(x)  dx^i]^2 + dx_idx^i  + dx^3 dx^3 \ , \ \  \
\
B_{it} = A_i (x) \ , \   \ \ \Phi=\Phi_0\ , }
\eqn\frf{     V_i=   e_0 A_i(x) = - {1\ov \sqrt {2\a'}} f \ep_{ij} x^j \ , \ \
\   \ \ e_0= \sqrt {2/\a'} \ , \ \ \ \ i,j=1,2 \ .      }
The magnetic field is  constant   in this
natural frame where the metric is stationary (it is  covariantly constant
in a general frame). The dilaton is  trivial and the
curvature and the antisymmetric tensor   vanish in the absence
of the magnetic field.

Let us first recall the form of the  actions of these models \rts\ (we shall
use  2d fermionic NSR
formulation). The $(1,1)$ extension \rts\ of the bosonic model of
\refs{\horts,\rts} is\foot{We shall use the following notation:
$\s_\pm = \s_0 \pm \s_1 \equiv \t \pm \s, \ \del_\pm = \ha (\del_0 \pm \del_1),
\ \s \in (0, \pi]  $.
The fermionic indices are coordinate ones  with  $ \l^\hv \equiv
\l^v + 2 A_i \l^i$.}
\eqn\onn{I_{(1,1)} ={  {1\ov \pi \a'} }
\int d^2 \s \big[ \del_+ u   \del_- v + \del_+ x_m   \del_-  x^m +
2A_i (x) \del_+ u \del_- x^i
  } $$
 +  \l^u_L \del_- \l^\hv_L  + \l_{Lm}\del_- \l^m_L  +  F_{ij} \del_- x^j
\l_{L}^u\l^i_L
 +   \l^\hv_R \del_+ \l_{R}^u + \l_{Rm}\del_+ \l^m_R
 -  F_{ij} \del_+  u  \l_{R}^i\l^j_R   \big] .  $$
Here $u\equiv y-t, \ v \equiv y +t$  whereas  $y\equiv y + 2\pi R$  is   the
internal KK coordinate. This model and its truncations discussed below
are `self-dual' with respect to
duality in $y$ direction (with $R\to \a'/R$).
$t, x^m $ $\ (m=1,2,3; \ i,j=1,2) $ are the 4-dimensional space-time
coordinates. The
isometry coordinate   $x^3$ is the  direction of  the constant magnetic field,
\eqn\maa{  A_i=-\ha F_{ij}x^j \ , \ \ \  \ \ \ F_{ij} = f \ep_{ij}  \ .     }
$\l^\m_L$ and $\l^\m_R$ are Majorana-Weyl fermions (we  omit  additional free
5 bosonic
and  5 left and 5  right fermionic  coordinates).
This model corresponds to an exact solution of type II superstring theory.
It preserves    space-time supersymmetry
\refs{\kallosh,\horts} and  the action
\onn\  has, in fact,  extended $(4,1)$
world-sheet supersymmetry.\foot{ This is not   surprising given that
 for  the non-compact $y$ the bosonic  model  admits  a plane-wave
interpretation
and  is equivalent
 to a non-semisimple WZW model \napwi.}

There are four `magnetic' heterotic models  which  are closely related to
this superstring model \onn, and, in particular, correspond to the same
space-time background \did,\frf: three  (`left-right' symmetric, `left'
and `right')  models  with KK embedding of the magnetic field and the `right'
model with the magnetic field embedded in the internal gauge sector.

The model  \onn\ can be  also    interpreted
 as   a heterotic \sm  \refs{\sen,\hulw}
corresponding to a `left-right symmetric'  heterotic  solution
obtained by the standard embedding of a  closed superstring solution into the
heterotic string theory.\foot{$\l_R^\m$ then play the role of the internal
fermions
  and $V^{ij}_u={\hat \omega}_{+ u}^{  ij }= - { F}^{ij} $ the role  of the
internal
gauge field.}  This solution   also
preserves  one half of maximal  space-time supersymmetry \kallosh\ and the
corresponding \sm is  $(4,1)$ supersymmetric.

  In addition,  there are two   non-trivial,   inequivalent   heterotic string
models
which are obtained by
 $(1,0)$  and $(0,1)$ supersymmetric truncations of \onn\
\horts.
Both  models  represent {  exact } heterotic string solutions  when combined
with
a free internal fermionic sector  (there is  no need
to introduce  non-vanishing   internal gauge field background \horts).  The
$(1,0)$
(but not the $(0,1)$ one) truncation   also  preserves  `one half'
of space-time supersymmetry ($N=2, D=4$) and has extended $(4,0)$
world-sheet supersymmetry \horts.
Omitting the additional free space-time and internal fermionic contributions
the  corresponding  actions can be written as follows:
\eqn\ono{I_{(1,0)} ={  {1\ov \pi \a'} }
\int d^2 \s \big[ \del_+ u   \del_- v +  \del_+ x_m   \del_-  x^m + 2A_i (x)
\del_+ u \del_- x^i
 } $$
 + \  \l^u_L \del_- \l^\hv_L + \l_{Lm}\del_- \l^m_L
 +  F_{ij} \del_- x^j \l_{L}^u\l^i_L \big]\  , $$
\eqn\ont{I^{\rm (kk)}_{(0,1)} ={  {1\ov \pi \a'} }
\int d^2 \s \big[ \del_+ u   \del_- v +  \del_+ x_m   \del_-  x^m  + 2A_i (x)
\del_+ u \del_- x^i
} $$
  + \   \l^\hv_R \del_+ \l_{R}^u + \l_{Rm}\del_+ \l^m_R
 -  F_{ij} \del_+  u  \l_{R}^i\l^j_R   \big]\ . $$
The above actions describe  string models with the
 magnetic field embedded in the
KK sector.
Starting with  the bosonic background \did,\frf\
one can construct the heterotic $\s$-model  where  the
 magnetic field    appears in the
internal  gauge  sector \rts\
\def\htt{ {\hat t} }
$$I^{\rm (int)}_{(0,1)} = {  {1\ov \pi \a'} }
\int d^2 \s \big[ -\del_+ t \del_- t    -  2A_i  \del_+ t  \del_- x^i  +
(\delta_{ij} - A_iA_j) \del_+ x^i   \del_- x^j    $$   $$
+ \del x^3 \bd x^3  -   \l^\htt_R \del_+ \l^\htt_R +   \l_{Ri}\del_+ \l^i_R
+  \l^{3}_R \del_+ \l^3_R   + F_{ij} \del_+  t    \l^i_R\l^j_R +  \ha F_{ij}
A_k  \del_+  x^k
\l^i_R\l^j_R
$$ \eqn\nnkp{
 +\   \bar \psi ( \del_- -  ie_0 A_i \del_- x^i) \psi
+  \ha i e_0  F_{ij}   \bar \psi \psi\l^i_R\l^j_R \ \big]
\   .    }
Here $
   e_0\equiv   {  \sqrt {2/\a'}}$ and $\l^\htt_R= \ha (\l^{\hat v}_R - \l^u_R)
$. The
 $\l^\m_R$ are  the four right Majorana-Weyl   fermions of the supersymmetric
sector
and $\psi$ is the left Weyl fermion of the internal sector which is coupled to
the magnetic field. The complete  anomaly-free
heterotic string model  is obtained  by adding to \nnkp\  extra free fields:
6  scalars and 6  right  and 30 left
 Majorana-Weyl fermions.

\def \y {\tilde y}

The model \nnkp\ admits also an alternative description with the coupling in
the internal sector
represented by a chiral boson.\foot{In general, the fermionic description
of the internal sector of  a heterotic string model  is more fundamental: it
can be replaced by a chiral bosonic
 one  only in  special cases.}
 This representation will be useful
for the solution of this heterotic string model, so let us discuss  it
in some detail.
The reason why we can give  a conformal and (on-shell) Lorentz-invariant
chiral boson description of this model is that the coupling  term
 $(A_i \del_-x^i  -\ha F_{ij} \l^i_R \l^j_R)
\del_+ y$ in the  closely related  model \ont\  is  linear in  the KK
coordinate  $y$
 and is chiral. As a result,
 $y$  can be consistently truncated
to its `chiral' part.\foot{As discussed in \rts,
\nnkp\ can be obtained  from \ont\
 (with decoupled   fermionic components  $\l^u_L, \l^\hv_L$) by
`fermionising' the  compact internal coordinate $y$ and dropping extra free
field terms in the action.}
  Following \refs{\flo,\tse} one can  describe this
coupling  by a chiral scalar  action  which is not manifestly Lorentz invariant
but
defines a  Lorentz-invariant
theory on-shell.
 Starting  with the bosonic $y$-dependent part of
\onn\ or \ont\
\eqn\doub{I(y,A_-)  = {  {1\ov \pi \a'} }
\int d^2 \s \big(  \del_+ y \del_- y    +   2A_-   \del_+ y ) \ ,
\ \ \     \     A_p\equiv A_i(x)  \del_p x^i  \ , }
and introducing the dual field $\td y$  one  finds
 the following `doubled' action
(see \tse\ and    section 2.4 in \rts)
\eqn\doube{I(y, \y,A_-)  = {  {1\ov 4\pi \a'} }
\int d^2 \s \big[  \del_0 y \del_1 \y    + \del_0 \y \del_1 y - \del_1 y \del_1
y
- \del_1 \y \del_1 \y } $$ + \    4 A_- (\del_1 y + \del_1 \y)
-  4A_-A_- \big] \ .   $$
Equation \doube\  is the same as the phase-space action with  momentum replaced
by
$\del_1\td  y$.
Integration over $\y$ leads back to \doube.
Written in terms of $y^\pm  = \ha (y \pm  \y)  $  \doube\ becomes
\eqn\dobe{I (y,\td y,A_-) = I(y^-) + I(y^+, A_-)\ ,
\ \    \ \ \ I(y^-) = - {  {1\ov \pi \a'} }
\int d^2 \s  \del_+ y^- \del_1 y^-    \ , }
\eqn\chir{ I(y^+, A_-) =  {  {1\ov \pi \a'} }
\int d^2 \s \big(
 \del_1 y^+ \del_- y^+  +   2 A_- \del_1 y^+
-  A_-A_- \big) \ .  }
The equations that follow (under proper boundary conditions) from \dobe\ and
\chir\
are
  $ \del_+ y^- =0, \ \ \del_- y^+ + A_-=0$. Since
$y^-$ is decoupled from the rest of the fields it can be consistently set equal
to zero.
Like  the original theory \doub, \doube , and the theory of the
free chiral scalar $y^-$,
 the theory defined by $I(y^+, A_-)$ is also   Lorentz-invariant on shell
(this can be easily checked  by computing  its stress energy tensor on the
equations of motion).
Since the equation of motion that follows from \doub\ is
$\del_+(\del_- y + A_-) =0$ the chiral truncation  corresponds to
choosing only the solutions which satisfy  $ \del_- y + A_-=0$
(note that  for generic $A_-$ such $y=y^+$  will depend on both $\t-\s$ and $\t
+\s$).
The action \chir\ can be rewritten also as\ \ ($D_p y\equiv  \del_p y + A_p $)
\eqn\chirr{ I(y^+, A_-) =  {  {1\ov \pi \a'} }
\int d^2 \s \big(
 D_1 y^+ D_- y^+      +  \ha \ep^{pq} A_p D_q y^+
  -  A_-A_+ \big) \ .   }
The equivalent form of the heterotic action  \nnkp\
with the bosonic representation of the internal sector is obtained by replacing
the fermionic $\psi$-terms, together with the  $A_+\hat A_-$-term
in \nnkp , by  $I(y^+, \hat A_-)$,
$$I^{\rm (int)}_{(0,1)} = {  {1\ov \pi \a'} }
\int d^2 \s \big[ - \del_+ t \del_- t     -  2\hat A_-  \del_+ t   +
 \del_+ x_m   \del_- x^m
 -   \l^\htt_R \del_+ \l^\htt_R +   \l_{Rm}\del_+ \l^m_R   $$
\eqn\bos{ +  \ \del_1 y^+ \del_- y^+   +  2  \hat A_- \del_1 y^+
-  \hat A_- \hat A_- \big] \ , \ \ \ \hat A_- \equiv A_i \del_-x^i  -\ha F_{ij}
\l^i_R \l^j_R \ . }
$y^+$ should be identified with one of the  coordinates $y^I_L$
of the  `left' 16-torus of the internal sector of the free heterotic string.
In general, the coupling to the 16  Abelian vector fields $A^I_\m$
 of Cartan subalgebra
is described by the action
\eqn\hett{I=
{  {1\ov \pi \a'} }
\int d^2 \s   \sum_{I=1}^{16} \big[ \del_1 y^I_L \del_- y^I_L   +  2  \hat
A_-^I (x)  \del_1 y^I_L
-  \hat A_- ^I\hat A_-^I \big] \  }
$$= {  {1\ov \pi \a'} }
\int d^2 \s \sum_{a,b=1}^{16}  g_{ab}  \big[ \del_1 y^a_L \del_- y^b_L   +  2
\hat A_-^a (x)  \del_1 y^b_L
-  \hat A_- ^a \hat A_-^b \big] \     ,  $$
where
\eqn\rer{ y^I_L\equiv y^I_L+ 2\pi R \sum_{a=1}^{16} n_a e^I_a ,
\  \ y^I_L=  \sum_{a=1}^{16} e^I_a y^a_L ,
 \  \ \hat A^I_-=  \sum_{a=1}^{16} e^I_a \hat A_-^a ,}
$$  R= \sqrt{\a'/2}\ , \ \
\   \  g_{ab} =  \sum_{I=1}^{16} e^I_a e^I_b \  ,  \ \ \  g_{aa}= 2 \ , $$
and $e^I_a$ are the  generators
of the even self-dual 16-lattice
($\Gamma_8\times \Gamma_8$ or $\Gamma_{16}$ \gross).

The two  `right' heterotic models \ont\ and \bos\ are closely related.
Indeed, \ont\ can be put into the form similar to \bos\
by first using that $u=y-t, \ v=y+t$ and then replacing the $y$-dependent part
of the action  (i.e. \doub\
with $A_-\to \hat A_-$)  by the equivalent form \doube :
$$I^{\rm (kk)}_{(0,1)} = {  {1\ov \pi \a'} }
\int d^2 \s \big[ - \del_+ t \del_- t     -  2\hat A_-  \del_+ t   +
 \del_+ x_m   \del_- x^m
 +  \l^{\hat v}_R \del_+ \l^u_R +   \l_{Rm}\del_+ \l^m_R   $$
\eqn\bot{ +\    \four( \del_0 y \del_1 \y    + \del_0 \y \del_1 y - \del_1 y
\del_1 y
- \del_1 \y \del_1 \y) +    \hat A_- (\del_1 y + \del_1 \y)
-  \hat A_- \hat A_- \big]  . }
In view of \dobe\ one can also trade  the $(y,\y)$-terms for the
$(y^-,y^+)$ ones. Then  it becomes  explicit  that \bos\ is
just the   $y^-=0, \  \l^{\hat y}=0 $ {\it truncation }
of \bot.

The  actions  \bos\  and  \bot \
are the special cases  of the action \ses\
of the $D=4$ heterotic string  compactified on a torus \narain.
Let $y^\a(\t,\s)$ be 28  fields that parametrize 28-torus
conjugate to an even self-dual lattice of signature (6,22).
The invariant metric of $O(6,22)$ can be chosen as
\def \b {\beta}
$$
{\cal L}_{\a\b} = \pmatrix{0 & I_6 & 0 \cr
I_6 & 0 & 0 \cr
0 & 0 & I_{16}\cr} .
$$
Introducing 28 Abelian vector fields $A^\a_\m$ and the matrix $M^{\a\b}$ of
moduli fields
$(M^T{\cal L} M={\cal L}, \ M^T=M)$ one finds that the (bosonic part of)
 $y^\a$-dependent terms in the manifestly $O(6,22)$  $T$-duality invariant
heterotic string action are given by  \refs{\ses,\tse}
$(D_p y^\a\equiv  \del_p y^\a + A_\m^\a \del_p x^\m $)\foot{$y^\a$
can be assumed to be compactified on circles of radii $R=\sqrt {\a'}$
with information about the specific torus being encoded in moduli.
Note that the upper block of ${\cal L}_{\a\b}$ is diagonalized
by setting $y={1\ov \sqrt 2} (y^1 + y^2), \
\y={1\ov \sqrt 2} (y^1 - y^2)$ with $y^{1,2}$   having the same normalization
as  $y^\a$.}
\eqn\chrr{ I  =  {  {1\ov 4 \pi \a'} }
\int d^2 \s \big[{\cal L}_{\a\b}
  D_0 y^\a   D_1 y^\b    -
({\cal L} M{\cal L})_{\a\b}  D_1 y^\a  D_1 y^\b
 +   \ep^{pq} A_p^\a {\cal L}_{\a\b}  D_q y^\b \big] .   }
The case of \doube,\bot\ corresponds to $y^\a=(y,\y), $
$M=I$, $A^\a_\m =A_\m$ , while that of \chir,\chirr,\bos\
corresponds to $y^\a=\sqrt 2 y^+, \  A^\a_\m =\sqrt 2 A_\m$.\foot{Note that the
action
\chrr\
in \ses\ differs from \doube,\chir,\chirr\ by the Lorentz-invariant
`counterterm' $A_-A_+$, which can be absorbed in the $x$-dependent part of the
action and must be present in \bos,\bot\ (for a discussion of this term in
connection with scheme dependence  see \rts).}

\newsec{Solution of the  heterotic string  models
 with  magnetic field in the Kaluza-Klein sector}
To determine the spectrum of the conformal models defined by \onn,\ono,\  \ont\
one may follow the same strategy  as used in the bosonic case in \rts.
The simplest model to solve is
\ono. In the light-cone gauge ($\l^u=0$)  the bosonic  and
fermionic variables essentially decouple and the solution reduces to that of
the bosonic model with trivial modifications due to the presence of the free
fermionic oscillators.
As a  consequence, one finds that  GSO projection \gso\ eliminates
tachyons  which were present in the bosonic case,  the spectrum is symmetric
between bosons and fermions and the partition function vanishes.
This conclusion is consistent with space-time supersymmetry of
 the corresponding
background \refs{\horts,\kallosh}.

 The resulting stability of this  heterotic  string
model  may be surprising
in view  of the conclusion \refs{\abo , \ferrara}\  that the
constant magnetic field background in the open superstring theory
is unstable
for certain values of the magnetic field. Indeed,  $F_{ij}=\const$ background
breaks space-time supersymmetry of this theory
and (as in the bosonic open string case)
 there are  tachyonic states in its  spectrum. The heterotic model \ono\ has
world-sheet supersymmetry
in the left (`charge') sector and,  therefore,  here
the presence of the magnetic  background does not spoil
the space-time supersymmetry.\foot{In particular,
$SO(32)$ or $E_8\times E_8$ gauge vector  bosons do not become
tachyonic   in the   heterotic model \ono\ because they are singlets under this
Kaluza-Klein  $U(1)$ group.}
The type II superstring model \onn\ inherits the space-time supersymmetry
of its `left part'  and is also stable (has no tachyons and  zero partition
function  after  the GSO projection).

It is the  model \ont\ that is the analogue of the open superstring model:
here the world-sheet supersymmetry is present only in the right sector
and as a result the space-time supersymmetry is broken (see also
\refs{\horts,\kallosh}).
As we shall show below, this model  has  indeed tachyonic instabilities
(in particular, the usual Yang-Mills ones).
It will  turn out that the  heterotic  $(0,1)$ model \nnkp,\bos\
 with the  gauge sector  embedding of the magnetic field  will have  similar
properties. Since its  solution  can be  obtained  from the solution of
\ont\ by a  `chiral truncation' (cf. \bos,\bot)  we shall first  consider the
latter  model  in
detail.


\subsec{Quantization and Virasoro conditions}

 Introducing
$x=x^1+ix^2, \ x^*=x^1-ix^2, \  \lambda_R =\lambda^1_R+i\lambda ^2_R ,\
\lambda ^*_R=\lambda^1_R-i\lambda ^2_R$ one can represent the  action
\ont,\maa\ in the form
(we omit additional free field terms and the  subscript $R$ on fermions)
\eqn\lagra{I^{\rm (kk)}_{(0,1)} ={  {1\ov \pi \a'} }
\int d^2 \s \big[ \del_+ u   \del_- v
+  \del_+ x   \del_-  x^*   +  \ha if \del_+ u ( x \del_-  x^*  -  x^*\del_- x)
}
$$
+ \  \lambda ^{\hat  v}\del_+ \lambda ^{u}+
  \lambda ^{*}\del_+ \lambda + i f\del_+ u  \l^*\l \big] \ . $$
The equations of motion are  given by
\eqn\equo{ \del_-\del_+ u=0 \ , \ \  \ \ \
\del_+ [\del_- v  +  \ha if ( x \del_-  x^*  -  x^*\del_- x+2\l^*\l )]=0\   ,
}
 \eqn\equatwo{\del_+\del_-x + if\del_+u\del_-x =0
  \ ,\ \ \ \ \ \  \del_+\del_-x^*-  if\del_+u\del_-x^* =0
 \ ,  }
\eqn\equu{ \del_+\l^{ u}=\del_+\l^{\hat v}=0\ ,\ \ \ \
 \del_+ \l  +   if \del_+u \l =0 \ , \ \  \ \ \  \del_+ \l^* - if \del_+u
\l^*=0
\  .  }
Since $u,\  \l^{ u}$ satisfy free equations, we can fix the remaining conformal
symmetry by  choosing the light-cone gauge \
$u = u_0+  p_+ \s_+ + p_-\s_-$, \ $ \l^{\hat u}=0$.
The general solution  of \equatwo\  is  then
\eqn\solus{
 x= e^{-if p_+\s_+} X\  , \  \ \ \  x^*= e^{if p_+\s_+}  X^* \
   ,  \  \ \ \ \   X= X_+ + X_- \ ,}
\eqn\solutwo{
v= v_+ + v_-  + \ha  if ( X^*_+ X_- -  X^*_- X_+ )\ ,
}
 \eqn\solutrhee {
\l=e^{-ifp_+\s_+}\eta _-\ ,\ \ \ \ \   \l^*=e^{ifp_+\s_+}\eta ^*_-
\ , }
where  the  subscripts $\pm$  indicate dependence on $\s_\pm =\t \pm \s$,
 i.e. $\ X_\pm = X_\pm (\s_\pm),$ etc.
The closed string periodicity conditions $\  x (\s + \pi,\tau) = x(\s,\tau)$,
$\ \l(\s+  \pi,\tau) = \pm \l(\s,\tau)$  are easily implemented by
setting
\def \X {{\cal X}}
\eqn\yplumin{
 X_+ = e^{ifp_+\s_+} \X_+ \ ,  \ \ \ \ \  X_-=  e^{-ifp_+\s_-} \X_- \ ,
\ \ \  \eta_- = e^{-ifp_+\s_-}\chi_- \ ,
}
where the new fields satisfy the standard  ``free-theory" boundary conditions,
$\X_\pm(\s,\t) = \X_\pm (\s+  \pi,\tau)$ and
$\chi_- (\s+  \pi,\tau) = \pm \chi_-(\s,\tau)$,   with the signs ``$\pm$"
corresponding to the Ramond (R)  and Neveu-Schwarz (NS)  sectors.
Thus
\eqn\fourie{ \X_+ =  i  \sqrt{\ha \a' } \sum_n \tilde a_n e^{-2in \s_+}   \ ,
 \ \     \X_- =  i  \sqrt{\ha \a' }
\sum_n a_n  e^{-2in \s_-}  \ , }
\eqn\fourram{ {\rm R:}\ \ \chi_-=\sqrt{2\a'}\sum_{n \in {\bf Z}} d_n
e^{-2in\s_-}\ ,
\ \ \ \ \ \ {\rm NS:}\ \ \chi_-=\sqrt{2\a'}\sum_{r\in {\bf Z}+1/2} c_r
e^{-2ir\s_-}\ .
}
The zero mode parts of the fields are $(u\equiv y-t, \ v\equiv y+t$)
\eqn\ze{
y_{\rm zero} = y_0   + 2L \s + k \t, \ \ \ \ \  \ t_{\rm zero}=t_0 + p\t
\ ,  \ \ }
\eqn\zer{  u_{\rm zero} =u = u_0 + p_+ \s_+ + p_-\s_- \ , \ \ \  p_\pm =
\pm L+ \ha ( k -p) \ , }
\eqn\zee{\ v_{\rm zero} =  v_0 + q_+ \s_+ + q_- \s_-\ , \ \
\ q_\pm  = \pm L + \ha ( k +p)\ .  }
If $y$ is compactified on a circle of radius $R$ then $L=Rw, \ w=0, \pm 1 , ...
$.
  The
stress tensor components corresponding to the model \lagra\  are  given by
 \eqn\tminus{
T_{--}=\del_- u \del_- v  +
 \del_- x \del_- x^*
+  \ha if \del_- u ( x \del_-  x^*  -  x^*\del_- x)
}
$$
+  \ i     \lambda ^{*} \del_-\lambda  -    f\del_- u  \l^*\l \ ,
$$
 \eqn\tplus{
T_{++}= \del_+ u \del_+ v  + \del_+ x \del_+ x^*
+ \ha if \del_+ u ( x \del_+  x^*  -  x^*\del_+ x)   \ . }
    The classical expressions for the Virasoro operators $L_0, \tilde L_0$ are,
in the R-sector,
\eqn\vira{
L_0^{\rm (R)}={1\ov 4\pi\a' } \int_0^\pi d\s\  T_{--} =   {p_-q_- \ov 4\a' }
 +\ha\sum_{n }  \big(n+\ha {fp_+}\big)\big(n+ f L\big) a_n^{*} a_{n}
}
$$+  \sum_n  (n+ fL) d_n^* d_n \ ,$$
 \eqn\soro{
\tilde L_0^{\rm (R)}={ 1\ov 4\pi\a' } \int_0^\pi d\s\  T_{++} =   {p_+q_+\ov
4\a'} + \ha
\sum_{n}  n\big( n-\ha  {fp_+} \big)       \td a_n^{*} \td a_{n} \ .
 }
The expressions in the NS sector are similar.
Using   \fourie\  we  obtain  from \lagra\  the canonical momentum  of $y $
\eqn\pphih{
 p_y={1\ov 2\pi\a'}\int_0^\pi d\s \big[ \del_0 y  +\ha if(x\del_-x^*-x^*\del_-
x + 2 \l^*\l ) \big]=\ha {\a '}\inv   k +  f { J}_R\ .
}
 ${ J}_R$ is    the  `right' part of the angular momentum
\eqn\momot{  J_R =- \ha \sum _n (n+\ha f p_+) a^*_n a_n   + K \ , }
\eqn\kqq{
 K^{\rm (NS)}=-  \sum_{r}^\infty c_{r}^* c_{r}
\ , \ \ \ \  \ \
 \ \ \  K^{\rm (R)}=-
  \sum_{n}^\infty d_{n}^* d_{n} \ .
 }
Since the background is stationary, the string    also has conserved energy
\eqn\epjr{
E=  \int_0^\pi d\s P_t=-{1\ov 2\pi\a'}\int_0^\pi d\s \big[ \del_0 t  +\ha
if(x\del_-x^*-x^*\del_- x + 2 \l^*\l)   \big] }
$$ = -\ha {\a '}\inv   p- f{ J}_R\  , $$
\eqn\quq{ p_+ = L + \a' (p_y + E) = \a'(Q_L + E)\ , \ \ \ \  \ \ \
Q_{L,R}\equiv  p_y \pm   {\a'}\inv L  \ .  }
Expressing  \vira\ and \soro\ in terms
of  $E,p_y,L$ (or $E, \ Q_{L,R}$) and oscillators, we obtain
\eqn\vvira{
L_0^{\rm (R)}= -\four \a'E^2+{\four} \a' Q_R^2
   +  \sum_{n }n  \big[ \ha (n+\ha {fp_+})  a_n^{*} a_{n} + d_n^*d_n \big]
-\ha fp_+ J_R \  }
$$=  -\four \a'E^2+\four \a' Q_R^2
   +  \sum_{n } (n+\ha {fp_+})\big[ \ha  (n+\ha {fp_+})  a_n^{*} a_{n} +
d_n^*d_n \big]
\ , $$
\eqn\ssoro{
\td L_0^{\rm (R)}= -\four \a'E^2+\four \a' Q_L^2
   +   \sum_{n }  \ha n (n-\ha {fp_+})  \td a_n^{*} \td a_{n}  -\ha fp_+J_R
\ .  }
We can now quantize the  theory in  a
standard way by promoting the Fourier modes to operators acting in  Fock space
and imposing  the canonical commutation relations. They imply the commutation
relations for the zero modes ($[y_0, p_y]=i$ so that the momentum eigenvalues
are  $p_y = mR\inv, \ m=0, \pm 1 , ... $) and
 \eqn\fff{[ a_n, a_m^{*}] =    2  (n+ \ha {fp_+} )\inv   \delta _{nm}\ , \ \ \
[ \td a_n, \td a_m^{*}] =    2  (n - \ha {fp_+} )\inv   \delta _{nm } \ . }
\eqn\fffddd{
\{ d_n,d_m^*\} = \delta_{nm}\ ,\ \ \ \{ c_r,c_s^*\} = \delta_{rs}\ .
}
Symmetrizing the classical expressions
for $L_0, \tilde L_0$ and $J_R$
we   then normal-order them and  use  the  generalized $\zeta$-function
prescription.\foot{In contrast to the bosonic case \rts\
here the $(fp_+)^2$ normal ordering terms cancel out
between bosons and fermions in \vvira\ and  do not appear  in \ssoro.
In the bosonic case the
modular invariance requires regularizing  the left and right sectors together
in a
symmetric way \rts\  (this is equivalent to using
$   \sum_{n=1}^\infty (n + a )= -{1\over 12}    + {1\over 2}   a (1-a)
$ in $L_0$ and adding the same normal-ordering constant in $\td L_0$). }
 In the R-sector the bosonic and fermionic  normal ordering constants in  $L_0$
 cancel out completely, i.e. one finds   only that $\td L_0\to \td L_0-1 $. In
the NS sector one obtains: \   $L_0\to $  $ L_0 -\ha $ $  + \four fp_+$ and
$\td L_0\to \td L_0-1+\four fp_+$.

To write down the resulting expressions for the Virasoro operators it is
convenient to  introduce  the creation and annihilation operators as follows
\eqn\ope{ [ b_{n\pm}, b_{m\pm }^{\dagger }] = \delta _{n m} \ ,\ \
 [\td b_{n\pm}, \td b_{m\pm }^{\dagger }] = \delta _{n m} \ ,\ \
[b_0,b_0^{\dag
}]=1 \ , \ \ [\td b_0,{\td b}_0^{\dagger }]=1 \ , }
\eqn\rightope{
 b_{n+}^{\dagger }= a_{-n} \omega_- \ ,\ \  b_{n+}= a_{-n}^* \omega_-\ ,\ \
 b_{n-}^{\dagger }= a_{n}^* \omega_+ \ ,\ \  b_{n-}= a_{n} \omega_+\ ,
}
\eqn\leftope{
 \td b_{n+}^{\dagger }= \td a_{-n} \omega_+ \ ,\ \  \td b_{n+}=\td a_{-n}^*
\omega_+\ ,\ \
\td b_{n-}^{\dagger }=\td a_{n}^* \omega_- \ ,\ \ \td b_{n-}=\td a_{n}
\omega_-\
,\
}
 \eqn\ttt{
b_0^{\dagger }=\ha \sqrt{fp_+} a_0^*  ,\  \ b_0=\ha \sqrt{fp_+}a_0
 , \  \ {\td b}_0^{\dagger }=\ha \sqrt{fp_+} \td a_0  ,\ \ \td b_0=\ha
\sqrt{fp_+}\td a_0^*
\  , }
where $\omega_\pm \equiv \sqrt {  \ha \big( n \pm \ha {fp_+} \big) }$, \
$n=1,2,...\ $.
The subscripts  $\pm$ correspond  to components with spin `up' and
`down' respectively. We have assumed that $0<fp_+<2$. For $fp_+>2$
or $fp_+<0$ the  creation/annihilation roles of some operators  change
  but the analysis
 remains essentially the same (see \rut \ for a detailed discussion of this
point). The Fock vacuum   obeys
also   $d_{-n}^*|0\rangle =d_{n}|0\rangle =0,\ n>0$ and
$\ c_{-r}^*|0\rangle =c_{ r}|0\rangle =0,\ r>0$.

Symmetrizing  and normal-ordering the classical expression  for $J_R$
\momot\  we get
\eqn\angulr{
{\hat  J}_R= - b^{\dagger }_0 b_0 -\ha    +\sum_{n=1}^\infty \big(  b^{\dag
}_{n+}b_{n+} - b^{\dagger }_{n-}  b_{n-} \big)+\hat K= J_R - \ha  \ , }
$$
 \hat  K^{\rm (NS)}=-  \sum_{r=1/2}^\infty (c_{r}^* c_{r} + c_{-r} c_{-r}^*)
 , \ \ \
 \ \hat  K^{\rm (R)}=-\ha [d_0^*,d_0] -
  \sum_{n=1}^\infty (d_{n}^* d_{n}+d_{-n}  d_{-n}^*)
 .  $$
The Virasoro operators \vvira, \ssoro\ should include also the contributions of
additional free degrees of freedom.
In the standard bosonic  description  of the heterotic string theory
\refs{\gross,\sche}
there are 16 left (internal sector) chiral bosons $y^I_L\  (I=1,..,16)$ (see
\hett, \rer)
  compactified on a torus corresponding  to the  even self-dual  16-lattice
\eqn\xxi{
y_L^I=y_0^I+\sqrt{2\a'} p_L^I\s_+ + i \sqrt{\ha \a' }\sum_{n\neq 0}
{1\ov n}\td \a_n^I e^{-2in\s_+}\ ,\ \ \  \ \ p_L^I=\sum_{a=1}^{16} n_a e^I_a\ ,
}
\eqn\percon{
y^I_L\equiv y^I_L+2\pi L^I\ ,\ \ \ \ \ \ L^I=\sqrt{\ha \a'}\sum_{a=1}^{16} n_a
e^I_a=\sqrt{\ha \a'} p_L^I\ .
}
 Including   also the contribution of the
remaining 5 free non-chiral  bosonic fields of the supersymmetric sector
($\a=5,...,9$)\foot{We are assuming that  the $D=10$ heterotic string is
compactified  on a 6-torus $T^6= S^1 \times T^5$  where $S^1$ is the
$y=x^4$-circle used to embed the Abelian magnetic field
and $T^5$ corresponds  to the additional free  coordinates.
 For simplicity,  we shall consider  only  the states which have zero winding
number in these 5 additional free dimensions. In
particular, various generalizations along the lines of \narain\ are
straightforward.}
we get from \vvira,\ssoro\
\def \N {\hat N}
\eqn\lzero{
{\hat  L}_0=
\four \a'(-E^2+p^2_\a +  Q_R^2) +\N_R
 - \ha \a '   f(Q_L +  E)  { \hat J}_R      \ ,
}
\eqn\lzerob{
{\hat  {\td L}}_0= \four   \a' ( -E^2+p^2_\a   + Q_L^2
)  + \ha {(p_L^I)}^2    + {\hat N_L}
- \ha   \a' f(Q_L+  E) {\hat  J}_R    \  ,  }
\eqn\hamiltoni{
\hat  H = {\hat  L}_0 +  {\hat  {\td L}}_0 = \ha   \a'[ -E^2 +  p_\a^2  +   \ha
(Q^2_L + Q^2_R)]  +\ha {(p_L^I)}^2 + \N_R+ {\hat N_L}  }
$$   - \
   \a' f(Q_L+    E)  {\hat  J}_R
   \ , $$
where
\eqn\coo{ \N_R= N_R -a \ , \ \ \  \N_L = N_L-1 \ , \ \ \ \
\ \  \   a^{\rm (R)} =0\ , \ \ \    a^{\rm (NS) } =\ha \ , }
\eqn\qqq{
Q_{L,R} = m R\inv  \pm  {\a'}\inv  wR \ ,
\ \ \ \ \ \four\a' (Q_L^2-Q^2_R) = mw \ ,
 }
and the free-theory
operators $N_L$ and $N_R$  are  (e.g.  in the Ramond sector):
\eqn\nnn{
 N^{\rm (R)}_R= \sum_{n=1}^\infty\big[ n ( b^{\dagger }_{n+}b_{n+}+ b^{\dagger
}_{n-}b_{n-}
+ b^{\dagger }_{n\a} b_{n\a} ) +  d^*_nd_n+d _{- n}d_{-n}^* + d _{-n\a}
d_{n\a}\big] \ ,\ }
\eqn\nn{
{ N}^{\rm (R)}_L= \sum_{n=1}^\infty\big[ n ( \td b^{\dagger }_{n+}\td b_{n+}+
\td b^{\dagger
}_{n-}\td b_{n-}+\td b^{\dagger }_{n\a} \td b_{n\a})  +
 \td \a _{-n}^I \td \a_{n}^I  \big]  \ .
}
The Virasoro conditions   are thus ${\hat  L}_0={\hat  {\td L}}_0=0$, i.e.,
\eqn\enn{ \hat H=0
  \ ,\ \ \  \ \     \N_R +  \four \a' Q^2_R= N_L -1  +  \four \a' Q_L^2  + \ha
(p_L^I)^2 .
}

Separating the spin part of the angular momentum,
$  {\hat J}_R = -  b^{\dagger }_{0}b_{0} -\ha  + S_R \to  -(l+ \ha)  + S_R$,
we obtain from \enn:
\eqn\ttt{
 E^2=p_\a^2 + Q_R^2+ 4{\a'}\inv \N_R  + f(Q_L+E)(2l+ 1)  - 2f(Q_L+E) S_R\
, \
}
where $l=0,1,2,...,$  is the  Landau
level.\foot{In the non-relativistic
 limit one finds the following expression  for
the  gyromagnetic factor of an arbitrary   physical state,
$g= 2(1+ {M\ov  Q_L} ){\langle S_R\rangle \ov \langle S\rangle}, $
which was discussed in \refs{\rts, \rut }\ in the context of the bosonic model.
 As  was pointed out in \rut, the presence of the term $O(M/Q_L)$ in this model
is accidental
and is due to the  non-vanishing antisymmetric tensor with strength
proportional to the magnetic field. The  universal expression
for the $g$-factor associated  to the intrinsic magnetic moment of the particle
in  heterotic string theory
is \rs:  $ g= 2 {\langle S_R\rangle \ov \langle S\rangle}$.  This expression
 was confirmed in \rut\ for a general class of exactly solvable  models
describing magnetic backgrounds.}

\subsec{Energy spectrum  and tachyonic instabilities}
The analysis similar  to the one carried out in the bosonic case \rts\ shows
that this model has tachyonic states in its
spectrum.
Indeed, the  resulting form of the Hamiltonian and level matching constraint
is very
similar to that in the bosonic case: the only differences are the presence of
the fermionic terms
in the operators $N_R,  J_R$, different
normal ordering constants and the  standard  free heterotic  string term
$(p_L^I)^2$
in the left Virasoro operator \lzerob.
Since the constraints \enn\ are expressed in terms of free
creation/annihilation operators  and are  diagonal in Fock space
the spectrum is found in the same way as in  the free heterotic string theory.

As follows from  \enn,   the equation for the energy spectrum  can be
represented  as\foot{For the purpose of identifying some tachyonic states in
the spectrum it is enough to consider only the states with vanishing momenta in
extra free directions, $p_\a=0$.}
 \eqn\ene {
 E^2=4{\a'}\inv \N_R  +    Q^2_R  - 2f(Q_L+E)\hat J _R \  ,
}
 \eqn\ener {
(E+ f\hat J _R )^2  =  4{\a'}\inv \N_R  +    (Q_L -  f \hat J _R)^2  + Q^2_R
-Q^2_L
  \  ,
}
or, equivalently,
 \eqn\enr {
(E+ f\hat J _R )^2  =  4{\a'}\inv (N_L -1)  +    (Q_L -  f \hat J _R)^2   \  .
}
The GSO projection \gso\  in the supersymmetric right sector
implies that  $\N_R$ can take  only non-negative integer values
($\N_R$ corresponds to  the number of states operator of  the light-cone
Green-Schwarz formulation).
As a result, there are no tachyons in the free ($f=0$ ) heterotic string
theory.  For a non-zero field $f$,
the energy levels of the free heterotic string
split according to the value of
the `right' contribution to the angular momentum  $J_R$ and the
`left' charge   $Q_L$.
As follows from \ene,\ener,\enr,  for $f\not=0$
there are states for which $E$  is complex.
This indicates  the presence of a tachyonic instability.\foot{This instability
is also reflected in the partition function  which
has infrared divergences  at those values of $f$ for which
the energy gets an imaginary component \rts.}
Equation \ener\  implies  that  the
tachyonic states  must have $\a'(Q^2_L -Q^2_R) = 4mw  > 0$,
i.e. belong to the winding sector.  From  \enr\  one learns  that  such states
necessarily must have $N_L=0$ ($N_L$ can take only values $0,1,2,...$).

One particular choice  of parameters and quantum numbers that leads to
tachyonic states  is
 \eqn\rrr{
 R=\sqrt {\a'}\ , \ \ \ m=w=1,2,...,  \ \  \ Q_R=0 \  ,  \ \  \ Q_L = 2m/\sqrt
{\a'}\ , \  \  \ p^I_L=0\ , } $$ \  \   \N_R = -1 + m^2\  , \ \ \ \ \ N_L=0 .
$$
Consider components with $\hat J_R>0$. It  follows from \enr\  that a given
state becomes tachyonic in the range  $f_{\rm cr (1)}>f>f_{\rm cr (2)}$,
\eqn\cri{ \sqrt{\a'} f_{\rm cr (1,2)}=  {2(m\pm 1)\ov  \hat J_R }\   . }
In particular, $m=1, \ \N_R=0 $ gives the standard charged vector (Yang-Mills)
instability  which appears   already for an infinitesimal  $f$.
For the states with  large $m$ and lying  on the leading Regge trajectory (with
maximal
$\hat J_R$, i.e. with zero orbital quantum number $l=0$ and maximal spin  at a
given level), $\N_R \simeq \hat J_R \sim m^2$,
we find that $\sqrt{\a'} f_{\rm cr} \simeq  2/m$.
Thus the higher the charge and spin of a given state, the smaller the
magnetic field   needed to make it tachyonic.

Also, for any given  arbitrarily small  $f$
there  exists an infinite number of tachyonic states
with large enough charges and spins.  For large $m$, these are states which in
the Regge diagram lie between the  parabolas,
$\hat J_R=c \sqrt{\hat N_R+1} \pm c, \ c=2/(\sqrt{\a'}f)$.
  Indeed, for fixed $f$    all states with
\eqn\parbl{
\ha\sqrt{\a' } f \hat J_R -1 < m<\ha\sqrt{\a' } f \hat J_R +1\ ,
\ \ \ m\geq m_0\ ,
}
where $m_0=\ha \big( c+ \sqrt{c^2-4(c-{3\ov 2}+a)} \big)\cong c$ , are
tachyonic. The   condition $m\geq m_0$ comes from the requirement $S_R\leq N_R
$.
 Since $m$ (with  $\hat J_R  $ satisfying \parbl ) can take
infinitely many possible values, there
are an infinite number of tachyons for any given magnetic field.
Similar  results were found in the bosonic string case \rts\ and
will apply also to the heterotic string model \bos\ considered in next section.
This pattern of tachyonic instabilities is different from the one found in the
open
superstring theory \ferrara. In particular, it reflects the fact that in
 closed string theory  there are states with arbitrarily large values of
charges.

Since  this discussion  applies  to both Neveu-Schwarz  and   Ramond  sectors,
there are also  an infinite
number of space-time fermions with an imaginary part in  the energy.
 This conclusion  is also different from what happens in the  open superstring
theory  where there are no tachyons in the R-sector \ferrara.
As expected, the  massless spin $\ha $ fermions  do not
 become tachyonic for $f\not=0$ (for them the contribution of the gyromagnetic
coupling cancels against  the energy of the zeroth Landau level).\foot{The
presence of higher spin fermionic states with complex energy does not
seem to be in conflict with
the standard  field-theory expectation that `tachyonic' fermions contradict
unitarity of the theory since here the
background  metric is non-static.}

 Now let us consider the model \ono. It  has supersymmetric left  and
non-supersymmetric  right  sector,
so that the  free-theory parts of  the Virasoro operators \lzero\ and \lzerob\
are interchanged  (with $p^I_R$ replacing $p_L^I$). The
 interaction ($f$-dependent)   term
is now purely bosonic (there is no fermionic contribution in $\hat J_R$).
As a result, the analogs of the conditions  \ene,\enn\ are
\eqn\enet {
 E^2=4{\a'}\inv \N_L  +    Q^2_L   - 2f(Q_L+E)\hat J_R \ , \
}
\eqn\conn{  N_R-1 +\ha (p_R^I)^2 = \N_L   +  mw     , \ \ \ \ \ \ \N_L\equiv
N_L-a \  .  }
The GSO projection here  applies to the left sector
implying that $\N_L=0, 1,2, ...$.
Now   $Q_L$
appears  both in  the interacting  {\it and } the free part of
the energy relation (cf. \ene\ and \enet) so that \enet\ can be put in
manifestly
non-negative form (cf. \ener)
 \eqn\enery {
(E+ f\hat J _R )^2  =  4{\a'}\inv \N_L  +    (Q_L -  f \hat J _R)^2  \  .
}
The expression for the energy spectrum in type II superstring model \onn\
is found by combining the above expressions and  again  is
manifestly non-negative.
The obvious  difference  with respect to the  heterotic model \ont\
is in the form of the  level matching constraint
(now $\N_R  = \N_L   +  mw $). Apart from the fact that $\hat J_R$  again
contains
the fermionic part,
the expression for $E^2$ is identically the same as
\enet\ or \enery.


Since the  magnetic field couples to the spin,  {\it a priori}
one expects that in any magnetic field background there will be a mass
 splitting between fermions and bosons, and hence   supersymmetry will be
necessarily broken.
One may wonder how the `left-right symmetric'  and   `left'  heterotic (and
type II superstring)   models
managed to preserve supersymmetry and hence avoid  tachyonic instabilities.
The reason is that here the magnetic field does not couple to the total spin,
but only to the right part of it, and the latter may happen to be the same for
fermions and bosons.\foot{Note that in heterotic string theory it is not
possible to   couple the
 magnetic field to the total spin.}  Both the  `left-right symmetric'  (or
type II
superstring)  and `left'  heterotic  models  still  have an  equal number of
bosons and fermions with
the same
$\hat J_R$ and, as a result, an equal number of bosons and fermions
at each level.
 The formal  mechanism responsible for avoiding tachyons in these models
is   GSO projection. It
 eliminated   not only ground state tachyon but also certain
higher level
 states of  the  free  bosonic string spectrum which otherwise would become
tachyonic in the presence of the  magnetic field.
For example,
the electrically charged  massless vector states  which appear
in the bosonic string compactified on a circle of radius $R=\sqrt{\a'}$,
and which become tachyonic in the presence of the magnetic field, are actually
projected out by GSO in the above two theories.

\newsec{Heterotic string model
with a magnetic field  in   the internal
gauge  sector}
Let us now describe the solution of  the `right' heterotic model \nnkp\ or
\bos\
where the magnetic field appears in the internal gauge symmetry  sector.
The two  `right' heterotic models \ont,\bot\  and \bos\ are closely related:
as discussed in Section 2, \bos\ is just a chiral truncation of \bot.
Given that the two  actions \bos\ and \bot\  are special cases of the  action
\chrr\
of the heterotic string  compactified on 28-torus,
which is manifestly   invariant under the $T$-duality group  $O(6,22)$,
it is natural to expect that the two
models have similar properties, in particular,
the  heterotic model \bos\ with the  gauge sector  (Cartan subalgebra)
embedding
of the Abelian magnetic field also  contains  tachyonic  states in its
spectrum.

The solution of the model  \bos\ is
found by repeating the discussion of the previous
section while  dropping the $y^-$ part of $y=y^+ (\t,\s) + y^-(\t-\s)$, i.e. by
choosing the special solution
$\del_- y^+ +  \hat A_- =0$ of the equation $\del_+(\del_+ y +  \hat A_- )=0$
in \equo,
\eqn\eqqq{ \del_- y^+  +  \ha if ( x \del_-  x^*  -  x^*\del_- x+2\l^*\l )=0 \
. }
Then  eqs. \solus--\zee\ still apply,  in particular,
\eqn\eee{ y^+_{\rm zero} = y^+_0   + 2L^+ \s + k^+ \t \ , }
%
%
Integrating  eq.\eqqq\  over $\s$
we now get (cf. \pphih,\momot)
\eqn\yrr{ k^+- 2L^+ + 2\a' fJ_R =0\ . }
The definition of the momentum \pphih\ is  also
modified ($\del_0 y^+$ does not appear in the interaction term in the action,
see  \chir),  but
the final expression  is still
  formally the same as in \pphih\ after  we use \yrr\
\eqn\mome{
 p_y^+={1\ov 2\pi\a'}\int_0^\pi d\s  \del_1 y^+ = {\a '}\inv   L^+
= \ha {\a'}\inv k^+ + fJ_R \   . }
As a result,
the `right' charge in \quq\   is now  equal to zero, i.e.
\eqn\cooo{Q_R=0\ , \ \ \ \ \ \ \  \ Q_L= 2{ \a'}\inv L^+ \ . }
The expressions for the Virasoro operators and the Hamiltonian  are  still
given  by \lzero--\hamiltoni\   with $Q_R=0$ and $Q_L^2$ being now  part of the
lattice momenta term $(p^I_L)^2$.
Indeed,  in this section
we  are  assuming that the  Abelian magnetic field is  embedded in the
internal gauge symmetry group by identifying  $y^+$  with
one of the  coordinates of the 16-torus, e.g. the first one,  $y^+= y^1_L$
(cf. \bos,\hett).
Then  (see \xxi,\percon)
\eqn\klk{L^+= L^1=\sqrt{\ha \a'} p^1_L  , \ \  \ \ Q_L= \sqrt {2{\a'}\inv}
p^1_L\   , \  } $$
\   \ \ (p^I_L)^2 = \ha \a'Q_L^2  + (p^{I'}_L)^2 , \ \ \ \ \
{I'}= 2,...,16  . $$
The  values of $L^+$ and   $Q_L$ are determined by the allowed values of
$p^1_L$
which depend on the choice of
one of the two possible integral even self-dual 16-lattices \gross.
The final expressions for the Virasoro operators are\foot{As in \hamiltoni\
 $p_\a$ are momenta of extra (here 6) dimensions which may be assumed, e.g., to
be compactified on a torus.}
\eqn\ero{
{\hat  L}_0=
\four \a'(-E^2+p^2_\a )+\N_R   - \ha \a '   f(Q_L +  E)  { \hat J}_R     \ ,
}
\eqn\rob{
{\hat  {\td L}}_0= \four   \a' (-E^2+p^2_\a )  + \ha {(p_L^I)}^2    + { \N_L}
- \ha   \a' f(Q_L+  E) {\hat  J}_R    \  ,  }
\eqn\oni{
\hat  H = \ha   \a'( -E^2 +  p_\a^2)  +\ha {(p_L^I)}^2+ \N_R+ { \N_L}  -
   \a'  f (Q_L +    E)  {\hat  J}_R
   \ ,  }
so that the analogues of the constraint \enn\ and the energy spectrum relation
\ene,\ener\ are
\eqn\eeq{   \N_R= \N_L    +\ha (p_L^I)^2\  , \ \ \ \ \   \ \N_L=N_L -1 \ , }
\eqn\eeee{
 E^2=4{\a'}\inv \N_R  + p_\a^2    - f(Q_L+E)\hat J _R \ .  }
  As in the case of bosonic string and heterotic string  with Kaluza-Klein
embedding \ont\
 discussed above,  the  expression for $E^2$
is not manifestly positive so that tachyonic instabilities are expected to
appear.

To determine the presence of
  states with complex energy  let us consider the simplest
 configuration with zero momenta in
6 extra dimensions  $\ p_\a=0$. Then  \eeq,\eeee\ imply
(cf. \ener,\enr)
\eqn\ere{
(E+ f\hat J _R )^2  =  4{\a'}\inv \N_R  +    (Q_L -  f \hat J _R)^2  -Q^2_L  \
,  }
 \eqn\eneri {
(E+ f\hat J _R )^2  =  4{\a'}\inv [N_L -1    + \ha   {(p_L^I)}^2 ]
 +    (Q_L -  f \hat J_R)^2  -Q^2_L \ . }
Note that \ere\  is the same as  the  condition \ener\ on the spectrum of
another `right' heterotic model \ont\
 in the special case of $Q_R=0$  \rrr\ discussed in the  previous section.
As in that model,  here
the tachyonic states may  also appear  only in the sector
with $N_L=0$.
   From the Virasoro constraint
 \eeq\ we   learn  that the condition $N_L=0$ (and the fact that after GSO
projection
$\hat N_R\geq 0$)
implies   that $(p^I_L)^2 \geq 2$.

In the simplest case of $(p^I_L)^2=2 , \ \N_R=0, \ N_L=0$,
which is analogous to the  $m=1$ case in \rrr\ and
corresponds  to the  charged vector bosons of the massless  heterotic string
level,
\ere\  reproduces  the standard   infinitesimal magnetic
instability of non-Abelian theory (considering $Q_L, f\hat J_R>0$, one has
 $(E+f\hat J_R)^2=-f\hat J_R (2Q_L-f\hat J_R )<0 $ for $f$ infinitesimal).
 Another  special choice that
 demonstrates the presence of tachyons at higher  string levels
is $p^{I'}_L=0 $ (cf. \klk).  Then  $  (p^I_L)^2 = \ha \a' Q^2_L$ and
\eeq,\eneri\ become  identically the same as the conditions \enn,\enr\
with   $Q_R=0, \ p^I_L=0$  \rrr, i.e. we get
\eqn\errr{  \N_R=  -1  + \four \a' Q_L^2 \  , \ \ \ \  N_L=0\ , \ \ \ \
\sqrt {\a'}Q_L= \sqrt 2 p^1_L \ , }
 \eqn\ennri {
(E+ f\hat J _R )^2  = -4{\a'}\inv  + (Q_L -  f \hat J_R)^2    \  . }
To find which  values  $p_L^1$  are actually possible
let us   express  $p_L^I$ in terms of the dual  generators,
$p_L^I=\sum_{a=1}^{16} m_a e^{*I}_a $, where $m_a$ are integers.
Then
$m_a=\sum_I p^I_L e^{I}_a=p^1_L e^1_a$,  i.e.
$p^1_L=m_1/e^1_1$.
In the basis of generators used in
 \gross\ the components  $e^1_a$ are either $\pm 1/2$ or $\pm 1$
(this applies to   both $\Gamma_8\times \Gamma_8$ and $\Gamma_{16}$
lattices).
 Typical charge configurations  thus give  $p^1_L=2m$, as can be
explicitly checked (note that $(p^I_L)^2$  must be even).
In this case the analogs of the conditions in \rrr\     are
 \eqn\iii{ p^1_L=2  m\  , \ \ \ \
  Q_L= 2\sqrt{{2\ov \a' }}
m\  ,   \ \ \  \ \N_R= -1 + 2m^2 \ , \ \  \ \ m=1,2,...\ . }
These  states become tachyonic for $f_{{\rm cr} (1)}>f>f_{{\rm cr} (2)}$
  with (cf. \cri)
\eqn\crii{ \sqrt{\a'} f_{{\rm cr}(1,2)}={2 (\sqrt{2}m\pm 1)\ov \hat J_R}\ .
}
The inequality analogous to \parbl\ shows that as  the bosonic model or  the
heterotic model with KK embedding,
the heterotic model with gauge sector embedding also has
an infinite number of tachyons  for any (e.g.  arbitrarily small)
value of the magnetic field strength $f$.

\newsec{Concluding remarks}
Generalizing the previous work  \rts\ we have shown  here  that,
as the  model of open superstrings,
the models of closed superstrings and heterotic strings  in constant magnetic
field are also exactly solvable.
The resulting structure of the string Hamiltonian
is very simple: it is given by the free-theory part
 plus the  gyromagnetic-type
interaction
term,  which is linear in the magnetic field strength (see \hamiltoni,\oni).

We have studied  in turn the two non-supersymmetric
heterotic string models  \ont\ and  \bos\
(with  world-sheet supersymmetry in the  `right'
sector), which correspond to the
 two possible ways to embed the Abelian magnetic field into
 heterotic string theory: (i) Kaluza-Klein embedding in the case of
 toroidal compactification from 10 to 4 dimensions,  and (ii)
embedding in the internal  gauge symmetry sector
of the 10-dimensional theory   (which can  be further compactified on some
manifold $M^6$).
While the second  case   is closer to realistic magnetic field backgrounds,
the two  models are    related
 (with the  latter  being  essentially a `truncation' of the former).  This is
not surprising given that
the internal gauge symmetry group of the heterotic string also
originates from a (chiral) compactification on a  special 16-torus \gross.
The two types of $U(1)$ gauge fields are indeed  particular  members of the
 set of 28 Abelian vector fields
which are present in  the  case of toroidal compactification of the  heterotic
string \narain.

The two  heterotic models
have similar properties.  In particular, both  exhibit  a
 tachyonic instability, i.e.
 contain states with complex energy in their  spectra.
The novel feature of the closed string theory
compared to the open string one is
 the presence of states with arbitrarily large values not only of masses and
spins but also of {\it charges}.
This  leads to a remarkable  closed string generalization
of the well-known magnetic instability of  non-Abelian gauge theory:
there exist an infinite number of closed string tachyonic states for any  value
of the magnetic field strength $f$.
 Since the
gyromagnetic coupling term  in  (mass)$^2$
 ($\sim M_0^2 - 2f Q_L \hat J_R+...$) is given by
 the product  of the magnetic field strength  $f$ with charge  $Q_L$
and angular  momentum (= spin $S_R$
  minus the Landau orbital momentum number), the states with
 the free string mass term  $M_0^2\sim m^2/\a'$,  spin $S_R \sim m^2$  and
charge $Q_L \sim m/\sqrt {\a'}$ will  become tachyonic
for $\sqrt {\a'}f \sim 1/m$.

 This instability  should apply to 10-dimensional
heterotic string as well as to any of its compactifications to 4 dimensions.
It  can be eliminated only if  massive states receive
Planck-mass  corrections  to their free-theory masses.
Thus in heterotic string theory there are directions in the space of possible
backgrounds
 along  which an infinitesimal
(supersymmetry breaking)  deformation produces
 infrared instabilities (which, being associated with both massless and massive
level states of the free theory,
  remain even after states of the massless level get small masses).

 It should be noted that since these infinitesimal instabilities  are  due
to states with large charges $Q$,  whose
  tree-level masses may receive  important  loop
corrections, it might disappear at the string loop level.
For example, if   we restrict  consideration to states
with $gQ<<1$, where $g$ is the string coupling, then the minimal  critical
magnetic field
will be of order $\sqrt{\a'} f \sim g$, i.e.  will no longer be  infinitesimal
(once massless level particles also  become  massive as a result of
 symmetry breaking).


In the context of the bosonic string theory   it was shown \rut\ that the same
pattern of instabilities appears also in a more general class of models
describing magnetic field configurations (in particular, with vanishing
antisymmetric tensor,
like  $a=1$ or $a=\sqrt 3$ dilatonic Melvin backgrounds).
 In these cases the  mass $M^2=E^2-p_\a^2$  is invariant with respect to the
residual Lorentz group  acting in directions orthogonal  to the
$(x_1,x_2)$-plane. We expect similar  tachyonic instabilities
to be present also in the   heterotic string versions of these models
(which   do not  preserve  space-time supersymmetry either).

\vskip 1cm
 \noindent {\bf Acknowledgements}

\noindent
We are grateful  to L. Alvarez-Gaum\'e,  C. Bachas,  M. Green, E. Kiritsis
 and  B. Schellekens for helpful discussions.
A.A.T.   acknowledges the  support
of PPARC, EC grant SC1$^*$-CT92-0789
and NATO grant CRG 940870.

\vfill\eject
  \listrefs
\vfill\eject
\end